%Paper: 9110007
%From: Peter Bouwknegt <BOUWKNEG%CERNVM.BITNET@CUNYVM.CUNY.EDU>
%Date: Wed, 02 Oct 91 10:23:30 SET

%%%%%%%%%%%%%%%%%%%%%%%%%%%%%%%%%%%%%%%%%%%%%%%%%%%%%%%%%%%%%%%%%%%%%%%
%%
%%                                                      CERN-TH.6267/91
%%
%%
%%      SOME ASPECTS OF FREE FIELD RESOLUTIONS IN 2D CFT
%%  WITH APPLICATION TO THE QUANTUM DRINFELD SOKOLOV REDUCTION
%%
%%
%%          P. Bouwknegt, J. McCarthy, and K. Pilch
%%
%%   (bouwkneg@cernvm , mccarthy@brandeis , pilch@physics.usc.edu)
%%
%%   to appear in the proceedings of "String and Symmetries 1991",
%%        Stony Brook, May 20-25, 1991.
%%
%%                   all macros included !
%%
%%%%%%%%%%%%%%%%%%%%%%%%%%%%%%%%%%%%%%%%%%%%%%%%%%%%%%%%%%%%%%%%%%%%%%%%

%%%%%%%%%%%%%%%%%%  tex macros for preprints, cm version %%%%%%%%%%%%%%
%                      P. Ginsparg
%                      last updated 9/89
%                      modified
%
%---------------------------------------------------------------------
\catcode`\@=11 % This allows us to modify PLAIN macros.

\newcount\yearltd\yearltd=\year\advance\yearltd by -1900

%
%      restores pagenumbers
%
%       use following instead of \Date on the preliminary draft,
%       puts date/time on each page in big mode, writes labels in margins

\def\draftmode{\message{ DRAFTMODE }\def\draftdate{{\rm preliminary draft:
\number\month/\number\day/\number\yearltd\ \ \hourmin}}%
\headline={\hfil\draftdate}\writelabels\baselineskip=20pt plus 2pt minus 2pt
 {\count255=\time\divide\count255 by 60 \xdef\hourmin{\number\count255}
  \multiply\count255 by-60\advance\count255 by\time
  \xdef\hourmin{\hourmin:\ifnum\count255<10 0\fi\the\count255}}}
%       use \nolabels to get rid of eqn, ref, and fig labels in draft mode
\def\nolabels{\def\wrlabel##1{}\def\eqlabel##1{}\def\reflabel##1{}}
\def\writelabels{\def\wrlabel##1{\leavevmode\vadjust{\rlap{\smash%
{\line{{\escapechar=` \hfill\rlap{\sevenrm\hskip.03in\string##1}}}}}}}%
\def\eqlabel##1{{\escapechar-1\rlap{\sevenrm\hskip.05in\string##1}}}%
\def\thlabel##1{{\escapechar-1\rlap{\sevenrm\hskip.05in\string##1}}}%
\def\reflabel##1{\noexpand\llap{\noexpand\sevenrm\string\string\string##1}}}
\nolabels
%
% tagged sec numbers
\global\newcount\secno \global\secno=0
\global\newcount\meqno \global\meqno=1
\global\newcount\mthno \global\mthno=1
\global\newcount\mexno \global\mexno=1
\global\newcount\mquno \global\mquno=1

\def\newsec#1{\global\advance\secno by1\message{(\the\secno. #1)}
\global\subsecno=0\xdef\secsym{\the\secno.}\global\meqno=1\global\mthno=1
\global\mexno=1\global\mquno=1
%\ifx\answ\bigans \vfill\eject \else \bigbreak\bigskip \fi  %if desired
\bigbreak\bigskip\noindent{\bf\the\secno. #1}\writetoca{{\secsym} {#1}}
\par\nobreak\medskip\nobreak}
\xdef\secsym{}
\global\newcount\subsecno \global\subsecno=0
\def\subsec#1{\global\advance\subsecno by1\message{(\secsym\the\subsecno. #1)}
\bigbreak\noindent{\it\secsym\the\subsecno. #1}\writetoca{\string\quad
{\secsym\the\subsecno.} {#1}}\par\nobreak\medskip\nobreak}
\def\appendix#1#2{\global\meqno=1\global\mthno=1\global\mexno=1%
\global\mquno=1
\global\subsecno=0
\xdef\secsym{\hbox{#1.}}
\bigbreak\bigskip\noindent{\bf Appendix #1. #2}\message{(#1. #2)}
\writetoca{Appendix {#1.} {#2}}\par\nobreak\medskip\nobreak}
%
%       \eqn\label{a+b=c}       gives displayed equation, numbered
%                               consecutively within sections.
%     \eqnn and \eqna define labels in advance (of eqalign?)
%
\def\eqnn#1{\xdef #1{(\secsym\the\meqno)}\writedef{#1\leftbracket#1}%
\global\advance\meqno by1\wrlabel#1}
\def\eqna#1{\xdef #1##1{\hbox{$(\secsym\the\meqno##1)$}}
\writedef{#1\numbersign1\leftbracket#1{\numbersign1}}%
\global\advance\meqno by1\wrlabel{#1$\{\}$}}
\def\eqn#1#2{\xdef #1{(\secsym\the\meqno)}\writedef{#1\leftbracket#1}%
\global\advance\meqno by1$$#2\eqno#1\eqlabel#1$$}
%
%          theorems and examples and questions
%
\def\thm#1{\xdef #1{\secsym\the\mthno}\writedef{#1\leftbracket#1}%
\global\advance\mthno by1\wrlabel#1}
\def\que#1{\xdef #1{\secsym\the\mquno}\writedef{#1\leftbracket#1}%
\global\advance\mquno by1\wrlabel#1}
\def\exm#1{\xdef #1{\secsym\the\mexno}\writedef{#1\leftbracket#1}%
\global\advance\mexno by1\wrlabel#1}
%
%                        footnotes
\newskip\footskip\footskip14pt plus 1pt minus 1pt %sets footnote baselineskip
\def\f@@t{\baselineskip\footskip\bgroup\aftergroup\@foot\let\next}
\setbox\strutbox=\hbox{\vrule height9.5pt depth4.5pt width0pt}
\global\newcount\ftno \global\ftno=0
\def\foot{\global\advance\ftno by1\footnote{$^{\the\ftno}$}}
%
%say \footend to put footnotes at end
%will cause problems if \ref used inside \foot, instead use \nref before
\newwrite\ftfile
\def\footend{\def\foot{\global\advance\ftno by1\chardef\wfile=\ftfile
$^{\the\ftno}$\ifnum\ftno=1\immediate\openout\ftfile=foots.tmp\fi%
\immediate\write\ftfile{\noexpand\smallskip%
\noexpand\item{f\the\ftno:\ }\pctsign}\findarg}%
\def\footatend{\vfill\eject\immediate\closeout\ftfile{\parindent=20pt
\centerline{\bf Footnotes}\nobreak\bigskip\input foots.tmp }}}
\def\footatend{}
%
%     \ref\label{text}
% generates a number, assigns it to \label, generates an entry.
% To list the refs on a separate page,  \listrefs
%
\global\newcount\refno \global\refno=1
\newwrite\rfile
\def\ref{\the\refno\nref}
\def\bref{\nref}
\def\nref#1{\xdef#1{\the\refno}\writedef{#1\leftbracket#1}%
\ifnum\refno=1\immediate\openout\rfile=refs.tmp\fi
\global\advance\refno by1\chardef\wfile=\rfile\immediate
\write\rfile{\noexpand\item{[#1]\ }\reflabel{#1\hskip.31in}\pctsign}\findarg}
%       horrible hack to sidestep tex \write limitation
\def\findarg#1#{\begingroup\obeylines\newlinechar=`\^^M\pass@rg}
{\obeylines\gdef\pass@rg#1{\writ@line\relax #1^^M\hbox{}^^M}%
\gdef\writ@line#1^^M{\expandafter\toks0\expandafter{\striprel@x #1}%
\edef\next{\the\toks0}\ifx\next\em@rk\let\next=\endgroup\else\ifx\next\empty%
\else\immediate\write\wfile{\the\toks0}\fi\let\next=\writ@line\fi\next\relax}}
\def\striprel@x#1{} \def\em@rk{\hbox{}}

\def\addref#1{\immediate\write\rfile{\noexpand\item{}#1}} %now unnecessary
\def\startrefs#1{\immediate\openout\rfile=refs.tmp\refno=#1}
\def\xref{\expandafter\xr@f}\def\xr@f[#1]{#1}
\def\refs#1{[\r@fs #1{\hbox{}}]}
\def\r@fs#1{\edef\next{#1}\ifx\next\em@rk\def\next{}\else
\ifx\next#1\xref #1\else#1\fi\let\next=\r@fs\fi\next}
%

%
% this is ugly, but moore insists
\newwrite\ffile\global\newcount\figno \global\figno=1
\def\fig{fig.~\the\figno\nfig}
\def\nfig#1{\xdef#1{fig.~\the\figno}%
\writedef{#1\leftbracket fig.\noexpand~\the\figno}%
\ifnum\figno=1\immediate\openout\ffile=figs.tmp\fi\chardef\wfile=\ffile%
\immediate\write\ffile{\noexpand\medskip\noexpand\item{Fig.\ \the\figno. }
\reflabel{#1\hskip.55in}\pctsign}\global\advance\figno by1\findarg}
\def\xfig{\expandafter\xf@g}\def\xf@g fig.\penalty\@M\ {}
\def\figs#1{figs.~\f@gs #1{\hbox{}}}
\def\f@gs#1{\edef\next{#1}\ifx\next\em@rk\def\next{}\else
\ifx\next#1\xfig #1\else#1\fi\let\next=\f@gs\fi\next}
\newwrite\lfile
{\escapechar-1\xdef\pctsign{\string\%}\xdef\leftbracket{\string\{}
\xdef\rightbracket{\string\}}\xdef\numbersign{\string\#}}

\def\writestop{\def\writestoppt{\immediate\write\lfile{\string\pageno%
\the\pageno\string\startrefs\leftbracket\the\refno\rightbracket%
\string\def\string\secsym\leftbracket\secsym\rightbracket%
\string\secno\the\secno\string\meqno\the\meqno}\immediate\closeout\lfile}}
\def\writestoppt{}\def\writedef#1{}
\def\seclab#1{\xdef #1{\the\secno}\writedef{#1\leftbracket#1}\wrlabel{#1=#1}}
\def\subseclab#1{\xdef #1{\secsym\the\subsecno}%
\writedef{#1\leftbracket#1}\wrlabel{#1=#1}}
\newwrite\tfile \def\writetoca#1{}
\def\leaderfill{\leaders\hbox to 1em{\hss.\hss}\hfill}
%       use this to write file with table of contents
\def\writetoc{\immediate\openout\tfile=toc.tmp
   \def\writetoca##1{{\edef\next{\write\tfile{\noindent ##1
   \string\leaderfill {\noexpand\number\pageno} \par}}\next}}}
%       and this lists table of contents on second pass

%
\catcode`\@=12 % at signs are no longer letters
%
%       Unpleasantness in calling in abstract and title fonts
\ifx\answ\bigans
 
 \font\titlei=cmmi10 scaled\magstep3
\font\titleis=cmmi7 scaled\magstep3 \font\titleiss=cmmi5 scaled\magstep3
\font\titlesy=cmsy10 scaled\magstep3 \font\titlesys=cmsy7 scaled\magstep3
\font\titlesyss=cmsy5 scaled\magstep3 
\else
 
 \font\titlei=cmmi10 scaled\magstep4
\font\titleis=cmmi7 scaled\magstep4 \font\titleiss=cmmi5 scaled\magstep4
\font\titlesy=cmsy10 scaled\magstep4 \font\titlesys=cmsy7 scaled\magstep4
\font\titlesyss=cmsy5 scaled\magstep4 
 
 \font\absi=cmmi10 scaled\magstep1
\font\absis=cmmi7 scaled\magstep1 \font\absiss=cmmi5 scaled\magstep1
\font\abssy=cmsy10 scaled\magstep1 \font\abssys=cmsy7 scaled\magstep1
\font\abssyss=cmsy5 scaled\magstep1 
\skewchar\absi='177 \skewchar\absis='177 \skewchar\absiss='177
\skewchar\abssy='60 \skewchar\abssys='60 \skewchar\abssyss='60
\fi
\skewchar\titlei='177 \skewchar\titleis='177 \skewchar\titleiss='177
\skewchar\titlesy='60 \skewchar\titlesys='60 \skewchar\titlesyss='60
\ifx\answ\bigans\else
 \fi
%

%
%---------------------------------------------------------------------
%%%%%%%%%%%%%%%%%%%%%%%%%%%%%%%%%%%%%%%%%%%%%%%%%%%%%%%%%%%%%%%%%%%%%%%%
%
%    Here our macros start
%
%%%%%%%%%%%%%%%%%%%%%%%%%%%%%%%%%%%%%%%%%%%%%%%%%%%%%%%%%%%%%%%%%%%%%%%%

\def\p{\partial}

\def\darr#1{\raise1.5ex\hbox{$\leftrightarrow$}\mkern-16.5mu #1}
\def\half{{\textstyle{1\over2}}} %puts a small half in a displayed eqn
%
%  Greek abbreviations (to be used within $....$)
%
\def\al{\alpha}
\def\be{\beta}
\def\ga{\gamma}  
\def\de{\delta}  \def\De{\Delta}
\def\ep{\epsilon}  
\def\ze{\zeta}
\def\et{\eta}

\def\ka{\kappa}
 \def\La{\Lambda}
\def\rh{\rho}
  
\def\ta{\tau}
  
\def\ph{\phi}  \def\Ph{\Phi}  
\def\ch{\chi}

%
%  boldface abbreviations  (to be used within $..$)
%

%

%
% Calligraphic abbreviations (to be used within $..$)
%

\def\cF{{\cal F}}
\def\cI{{\cal I}}

\def\cO{{\cal O}}
\def\cU{{\cal U}}

\def\cW{{\cal W}}

%
%
%  definition of \underhook{...}
%
\def\lefthook{{\vrule height5pt width0.4pt depth0pt}}
\def\righthook{{\vrule height5pt width0.4pt depth0pt}}
\def\leftrighthookfill{$\mathsurround=0pt \mathord\lefthook
     \hrulefill\mathord\righthook$}
\def\underhook#1{\vtop{\ialign{##\crcr$\hfil\displaystyle{#1}\hfil$\crcr
      \noalign{\kern-1pt\nointerlineskip\vskip2pt}
      \leftrighthookfill\crcr}}}
%
%  other abbreviations
%
\def\KMa{Kac-Moody algebra}
\def\hwm{highest weight module}

\def\cfts{conformal field theories}
\def\dCo{{\rm h}^{\vee}}

\def\ie{{\it i.e.\ }}
\def\eg{{\it e.g.\ }}

\def\ZZ{Z\!\!\!Z}               %%%%%%%%%%%%%
                 %
                 % produces reasonably looking math. Z,N,etc.
\def\CC{I\!\!\!\!C}             %
\def\QQ{I\!\!\!\!Q}             %%%%%%%%%%%%%
\def\bfg{{\bf g}}

\def\bfnm{{\bf n}_-}
\def\bfnp{{\bf n}_+}

\def\dz#1{{\p\over \p z_{#1} }}
\def\bs{\bigskip}

    % Use within $..$

\def\nest#1{[\![\, #1\, ]\!]}

\def\alp{\al_+}

%%%
%%%
%%% some macros for making commutative diagrams
%%%

\def\mape#1{\smash{\mathop{\longrightarrow}\limits^{#1}}}

%%%%%%%%%
% Krzysztof's definitions
%

\def\state#1{|#1\rangle}

%%%%%%%%%%%%%%%%%%%%%%%%%%%%%%%%%%%%%%%%%%%%%%%%%%%%%%%%%%%%%%%%%%%%%%%%%%
%%%%
%%%%   END MACRO FILE
%%%%
%%%%%%%%%%%%%%%%%%%%%%%%%%%%%%%%%%%%%%%%%%%%%%%%%%%%%%%%%%%%%%%%%%%%%%%%%%

\magnification=1200

%%%%%%%%%%%%%%%%%%%%%%%%%%%%%%%%%%%%%%%%%%%%%%%%%%%%%
%%
%%            SOME MORE MACROS
%%
\def\Walg{$\cW$-algebra}
\def\Wg{\cW[\bfg]}
\def\wW{\widehat{W}}
\def\wDe{\widehat{\De}}
\def\wLa{\widehat{\La}}
\def\wrh{\widehat{\rh}}
\def\wal{\widehat{\al}}
\def\wbe{\widehat{\be}}
\def\whg{\widehat{\bf g}}
\def\whn{\widehat{\bf n}}
\def\cC{{\cal C}}
\def\cS{{\cal S}}
\def\Lap{\La^{(+)}}
\def\Lam{\La^{(-)}}
\def\Tr{{\rm Tr\,}}
\def\deg{{\rm deg}}

\def\Heca#1{H_Q^{#1}(\whg_k,\widehat{\bf n},\ch_{DS})}
\def\Hecm#1#2{H_Q^{#1}(\widehat{\bf n},\ch_{DS},#2)}
\def\fancyhatg{{\rm Hom\,}_{{\cal U}( {\hbox{{\hbox{$\scriptstyle\bf g$}}
 \kern-.8em\lower.5ex\hbox{$\scriptstyle\widehat{\phantom{\bf g}}$}}})}}
\def\bigoneoverp{{\hbox{$\scriptstyle 1\over 2pp'$}}}
%%
%%               END MACROS
%%
%%%%%%%%%%%%%%%%%%%%%%%%%%%%%%%%%%%%%%%%%%%%%%%%%%%%%
%%%%%%%%%%%%%%%%%%%%%%%%%%%%%%%%%%%%%%%%%%%%%%%%%%%%%
%%
%%               REFERENCES
%%
\def\NP{Nucl. Phys.\ }
\def\PL{Phys. Lett.\ }
\def\CMP{Comm. Math. Phys.\ }

\def\IJMP{Int. Journ. Mod. Phys.\ }

\def\FFrd{\FFkniz}
\bref\BMPa{
P. Bouwknegt, J. McCarthy and K. Pilch, Prog. Th. Phys. Sup.
{\bf 102} (1990) 67.}
\bref\FLa{
V.A. Fateev and S.L. Lukyanov, \IJMP {\bf A3} (1988) 507.}
\bref\FL{
V.A. Fateev and S.L. Lukyanov, {\it Additional symmetries
and exactly soluble models in two dimensional conformal field
theory}, Kiev lecture notes, 1988.}
\bref\BO{
M. Bershadsky and H. Ooguri, \CMP {\bf 126} (1989) 49.}
\bref\Fi{
J.M. Figueroa-O'Farrill, \NP {\bf B343} (1990) 450.}
\bref\FFrb{
B.L. Feigin and E.V. Frenkel, \PL {\bf 246B} (1990) 75.}
\bref\Fr{
E.V. Frenkel, {\it Affine \KMa s at the critical level and quantum
Drin\-feld-Sokolov reduction}, Harvard University thesis, May 1991.}
\bref\FKW{
E.V. Frenkel, V.G. Kac and M. Wakimoto, in preparation and private
communication.}
\bref\KWa{
V.G. Kac and M. Wakimoto, Proc. Natl. Acad. Sci. USA {\bf 85}
(1988) 4956.}
\bref\KWb{
V.G. Kac and M. Wakimoto, Adv. Ser. Math. Phys. {\bf 7} (1989) 138.}
\bref\KWc{
V.G. Kac and M. Wakimoto, {\it Branching functions for winding
subalgebras and tensor products}, MIT preprint, 1990.}
\bref\Wa{
M. Wakimoto, \CMP {\bf 104} (1986) 605.}
\bref\FFra{
B.L. Feigin and E.V. Frenkel, Usp. Mat. Nauk {\bf 43} (no.5) (1988) 227.}
\bref\BMPb{
P. Bouwknegt, J. McCarthy and K. Pilch, \CMP {\bf 131} (1990) 125.}
\bref\FFkniz{
B.L. Feigin and E.V. Frenkel,
in {\it Physics and Mathematics of Strings}, eds. L.~Brink {\it et. al.}
(World Scientific, Singapore, 1990).}
\bref\DM{
P. Deligne and G.D. Mostow, Publ. IHES {\bf 63} (1986) 5.}
\bref\TK{
A. Tsuchiya and Y. Kanie, Publ. RIMS {\bf 22} (1986) 259.}
\bref\FF{
B.L. Feigin and D.B. Fuchs, in {\it Representations of infinite-dimensional
Lie groups and Lie algebras} (Gordon and Breach, New York, 1989).}
\bref\KK{
V.G. Kac and D.A. Kazhdan, Adv. Math. {\bf 34} (1979) 97.}
\bref\FFrc{
B.L. Feigin and E.V. Frenkel, \CMP {\bf 128} (1990) 161.}
\bref\SVa{
V.V. Schechtman and A.N. Varchenko, {\it Arrangements of hyperplanes and
Lie algebra homology}, Moscow preprint, Aug. 1990.}
\bref\SVb{
V.V. Schechtman and A.N. Varchenko, {\it Quantum groups and homology of
local systems}, Princeton preprint, Nov. 1990.}
\bref\FW{
G. Felder and C. Wieczerkowski, \CMP {\bf 138} (1991) 583.}
\bref\GS{
C. Gomez and G. Sierra, \PL {\bf 240B} (1990) 149.}
\bref\RRR{
C. Ramirez, H. Ruegg and M. Ruiz-Altaba, \PL {\bf 247B} (1990) 499.}
\bref\Ro{
M. Rosso, \CMP {\bf 124} (1989) 307.}
\bref\BF{
D. Bernard and G. Felder, \CMP {\bf 127} (1990) 145.}
\bref\Wat{
G. Watts, \NP {\bf B361} (1991) 311.}
\bref\It{
K. Ito, {\it Quantum Hamiltonian reduction and $\cW B$ algebra},
YITP/K-885.}
\bref\DGK{
V.V. Deodhar, O. Gabber and V.G. Kac, Adv. Math. {\bf 45} (1982) 92.}
\bref\RW{
A. Rocha-Caridi and N. Wallach, Math. Z. {\bf 180} (1982) 151.}
\bref\BBSS{
A. Bais, P. Bouwknegt, K. Schoutens and M. Surridge, \NP {\bf B304}
(1988) 348, 371.}
\bref\Fe{
G. Felder, \NP {\bf B317} (1989) 215; erratum, {\it ibid.} {\bf B324} (1989)
548.}
\bref\BMPc{
P. Bouwknegt, J. McCarthy and K. Pilch, \NP {\bf B352} (1991) 139.}

%%
%%                 END REFERENCES
%%
%%%%%%%%%%%%%%%%%%%%%%%%%%%%%%%%%%%%%%%%%%%%%%%%%%%%%%%%%%%%%%%%%%%%%%%
%%
%%
%%                   TITLEPAGE
%%
%%
\nopagenumbers\pageno=0
\line{\hfill{CERN-TH.6267/91}}
\bigskip
\centerline{{\bf SOME ASPECTS OF FREE FIELD RESOLUTIONS IN 2D CFT}}
\centerline{{\bf WITH APPLICATION TO THE QUANTUM}}
\centerline{{\bf DRINFELD-SOKOLOV REDUCTION}\footnote{$^*$}{
To appear in the proceedings of ``Strings and Symmetries 1991,''
Stony Brook, May 20-25, 1991.} }
\bigskip\bigskip
\centerline{Peter Bouwknegt}
\smallskip
\centerline{{\it CERN - Theory Division}}
\centerline{{\it CH-1211 Geneva 23}}
\centerline{{\it Switzerland}}
\bigskip
\centerline{Jim McCarthy}
\smallskip
\centerline{{\it Department of Physics}}
\centerline{{\it Brandeis University}}
\centerline{{\it Waltham, MA 02254}}
\bigskip
\centerline{{ Krzysztof Pilch}}
\smallskip
\centerline{{\it Department of Physics}}
\centerline{{\it University of Southern California}}
\centerline{{\it Los Angeles, CA 90089-0484}}
\bigskip
\bigskip
\centerline{{\bf Abstract}}\smallskip
We review some aspects of the free field approach to
two-dimensional \cfts.
Specifically, we discuss the construction of free field resolutions for
the integrable \hwm s of untwisted affine \KMa s, and extend the
construction to a certain class of admissible \hwm s. Using these, we
construct resolutions of the completely degenerate \hwm s of \Walg s by
means of the quantum Drinfeld-Sokolov reduction. As a corollary we derive
character formulae for these degenerate \hwm s.

\bs

\vfil
\line{ CERN-TH.6267/91   \hfill }
\line{ BRX TH-325 \hfill}
\line{ USC-91/29 \hfill September 1991}

\eject
%%
%%             END TITLEPAGE
%%
%%%%%%%%%%%%%%%%%%%%%%%%%%%%%%%%%%%%%%%%%%%%%%%%%%%%%%%%%%%%%%%%%

%%%%%%%%%%%%%%%%%% acknowledgement %%%%%%%%%%%%%%%%%%%%%%%%%%%%%%%%%
\def\acknow{
\newsec{Acknowledgements}

P.B.\ would like to thank E.\ Frenkel for discussions.
The work of J.M.\ was supported by the NSF Grant \#PHY-88-04561 and the
work of K.P.\ was supported in part by funds provided by the
DOE Contract \#DE-FG03-84ER-40168 and by the USC Faculty Research
and Innovation Fund. Finally, we all thank the organizers for an enjoyable
and stimulating conference.
}
%%
%%                     PAPER
%%
%%
\footline={\hss\tenrm\folio\hss}
\baselineskip=1.2\baselineskip

\newsec{Introduction}

Free field techniques have been widely used in the study of
two-dimensional conformal field theories.
Realizations of the chiral algebra by free fields, resolutions which project
onto its irreducible representations from the Fock space modules,
and chiral vertex operators intertwining between resolutions, form the
tools of this approach (see the review [\BMPa], and references therein).

In this paper we intend to highlight several points which have not been
emphasized in the literature, and refer the reader to [\BMPa] for a
more complete discussion of the basics.  The main theme will be the
application to quantum Drinfeld-Sokolov reduction, which allows one to
investigate virtually all properties of \Walg s using corresponding
properties of \KMa s.  Our presentation of this application is very
much inspired by, in particular, references [\FLa--\FKW] in which most
of the results can be found.

The paper is organized as follows. In Section 2 we review the
free field realizations of affine \KMa s, and their associated
screening operators.  The construction of intertwining operators
between free field Fock spaces, and how these can be used to build a
complex of free field Fock spaces yielding a resolution of an irreducible
highest weight module, is explained in Section 3. Here we will not restrict
ourselves to integrable highest weights -- relevant for the corresponding
WZW-model -- but also treat a class of admissible weights introduced by Kac
and Wakimoto [\KWa--\KWc]. This is the class of weights relevant for the
quantum Drinfeld-Sokolov reduction since they lead to completely degenerate
highest weight modules for the corresponding \Walg. This will be discussed
in Section 4. Finally, in Section 5, we apply the results of Section 4
to derive character formulae for the completely degenerate \Walg\
modules, and compare them to the characters obtained in the coset model
approach to \Walg s.

For basic notations used throughout this paper the reader is invited to
consult Section 1.1 in [\BMPa].

\newsec{Free field realizations of affine \KMa s}

In this section we briefly outline the construction of realizations of
the (untwisted)
affine \KMa\ $\whg$ on free field Fock spaces. The first example of
such a realization was constructed by Wakimoto for
$\widehat{sl(2)}$ [\Wa]. In
general one proceeds as follows [\FFra,\BMPb]. Let $G$ be the (complex)
group corresponding to $\bfg$ and  let $B_-$ be a Borel subgroup. A
character $\ch_\La\,:\,B_-\to\CC^*$ defines a
holomorphic line bundle over the flag
manifold $B_-\backslash G$. The group $G$, and thus also $\bfg$, act on
(local) sections of this line bundle. Upon introducing complex
coordinates $z_\al\,,\al\in\De_+$, on a maximal cell of $B_-\backslash G$
this becomes a realization of $\bfg$ in terms of linear differential
operators. Restricting this realization to polynomials in $z_\al$
provides us with the ``analogue'' free field realization for finite
dimensional Lie algebras.\foot{See [\BMPa] for an
algebraic equivalent of this construction.}
This realization can be lifted to a
realization of the affine \KMa\ $\whg$ at level $k$ [\FFra].

Specifically, introduce a set of bosonic first order fields
$(\be^\al(z),\ga^\al(z))\,,\al\in \De_+$, of conformal dimension $(1,0)$
(corresponding to $\dz{\al}$ and $z_\al$, respectively), and
a set of ${\rm rank\,}\bfg=\ell$ scalar fields $\ph^i(z)$, with operator
product expansions
$\ga^\al(z)\be^\al(w)  \sim \de^{\al\al'}/(z-w)$ and
$\ph^i(z)\ph^j(w)  \sim -\de^{ij} \ln(z-w)$.
Introduce furthermore the Fock spaces $\cF_\La = \cF^\ph_\La\otimes
\cF^{\be\ga}$ that are freely generated from the vacuum $\state{\La}$
by the oscillators
$\be_n^\al\,,a^i_n$ for $n<0$, $\ga^\al_n$ for $n\leq0$, and labeled
by the momentum zero mode of the scalar field
$p^i\state{\La} = \alp\La^i\state{\La}$.
Here $\al_+^{-1} = \sqrt{k+\dCo}$, and we will also use the notation
$\al_-$, where $\al_+\al_-=-1$.

The following expressions for the
Chevalley generators define a realization of $\whg_k$ on $\cF_\La$ of
highest weight $\La$
\eqn\PBBe{
\eqalign{
e_i(z) & = \be^{\al_i} (z) + \ldots\,, \cr
h_i(z) & = -\al_-(\al_i^\vee\cdot i\p\ph(z)) + \sum_{\al\in\De_+}
(\al,\al_i^\vee) : \ga^\al(z)\be^\al(z):\,, \cr
f_i(z) & = \al_- \ga^{\al_i}(z) ( \al_i^\vee\cdot i\p\ph(z)) +
\ldots\,. \cr}
}
Suppose we introduce a degree by $\deg(\be^\al)=-1\,,\deg(\ga^\al)=1\,,
\deg(\ph^i)=0$. Then in $e_i(z)$ the dots stand for terms of $\deg\geq0$
in terms of $\be\ga$-fields only. The term of $\deg=-1$ is basis
independent while the higher degree terms depend on the particular choice
of basis. The dots in $f_i(z)$ stand for terms of $\deg\leq1$ in terms
of $\be\ga$-fields only, and  are basis dependent (unfortunately even the
terms of highest degree).
They are however
completely fixed once a particular choice for $e_i(z)$ has been made.

The Virasoro algebra acts on $\cF_\La$ by means of the Sugawara
construction, which in the realization \PBBe\ takes the form
\eqn\PBBf{
T(z) = -\half :\p\ph(z)\cdot\p\ph(z): - \alp \rh\cdot i\p^2\ph(z) -
\sum_{\al\in\De_+} :\be^\al(z)\p\ga^\al(z):\,.
}
The $L_0$-eigenvalue of the Fock space vacuum $\state{\La}$,
\ie its conformal dimension, is given by
$h_\La = \half \alp^2 (\La,\La+2\rh)$.

Finally, we need operators acting between Fock spaces labeled by
{\it different} weights $\La$ -- the so-called screening operators --
which will be used to build operators that intertwine with the action of
$\whg_k$. In the finite-dimensional analogue problem they arise from
the (left)
action of $\bfnp$ on $B_-\backslash G$, giving differential operators
which clearly commute with $\bfnp$ in the realization. Lifting these
to $\whg_k$ gives operators of the form
\eqn\PBBh{
s_i(z) = -\left( \be^{\al_i}(z) + \ldots \right)
e^{-i\alp\al_i\cdot\ph(z)}\,,\qquad i=1,\ldots,\ell\,,
}
where the dots stand for terms of $\deg\geq0$ in terms of
$\be\ga$-fields only.
The operator products of $s_i(z)$ with $e_i(z)$ and $h_i(z)$ are
regular, while
\eqn\PBBi{
f_i(z) s_j(w) \sim -\de_{ij} {2(k+\dCo)\over (\al_i,\al_i)}
\p_w \left(
{1\over z-w} e^{-i\alp\al_i\cdot\ph(w)} \right)\,.
}
For $\widehat {\bf g}_k=\widehat {sl(n)}_k$,  \PBBi\ can  be verified
directly, using the explicit realization of the currents $f_i(z)$ and
$s_i(z)$ [\FFkniz,\BMPb]. In the general case this result, conjectured
in [\BMPa], can be proved by observing that an algebraic field
redefinition reduces the computation to an $\widehat {sl(2)}$
subalgebra [\Fr].

\newsec{Intertwiners and resolutions}

The operator product expansions of  $s_i(z)$ with $e_i(z)$,  $h_i(z)$
and $f_i(z)$  discussed in the previous section provide the starting point
for constructing intertwiners, which are mappings between Fock spaces
that commute with the action of the current algebra. In particular
\PBBi\ suggests we consider the (formal) operators
$Q_{\La,\La'}(\cC):\cF_\La \rightarrow  \cF_{\La'}$,
\eqn\PBCa{
Q_{\La,\La'}(\cC) = \nest{s_{i_1}\ldots s_{i_n}}
= \int_{\cC} dz_1\ldots dz_n\ s_{i_1}(z_1)\ldots s_{i_n}(z_n)\,,
}
where  $\La'=\La-\be$ (note that $\be=\sum_j\al_j$ is
a sum of {\it positive} roots), and
$\cC$ is a suitably closed  multi-contour as discussed below.
More precisely, when acting on a vector $v\in \cF_\La$,
\eqn\PBCb{
\eqalign{
Q_{\La,\La'}(\cC )v = \sum_{l_1,\ldots,l_n\in\ZZ}
\int_{\cC} dz_1\ldots dz_n\ &\prod_{k<l} (z_k-z_l)^{
\alp^2(\al_{i_k},\al_{i_l})} \prod_k z_k^{-\alp^2(\La,\al_{i_k})} \cr
& \times \prod_k z_k^{-l_k-1} :\widehat{s}_{i_1,l_1}\ldots
\widehat{s}_{i_n,l_n}:\,T_{\La}^{\La'}v\,. \cr}
}
Here, $\widehat{s}_i$ denotes the screening operator with the scalar zero
modes removed and $\widehat{s}_i(z) = \sum_{n\in\ZZ} \widehat{s}_{i,n}
z^{-n-1}$.  The translation operator $T_{\La}^{\La'}$ maps
$\state{\La}$ to $\state{\La'}$.  Finally, $\cC$ is a contour in
$H_n(M,\cS)$ where $M= \{ (z_1,\ldots,z_n) \in\CC^n|z_i\neq0\,,z_i\neq
z_j (i\neq j) \}$ and $\cS$ is the (rank one) local system associated
to the multivalued integrand of \PBCb\ [\DM].  In other words $\cC $
must be a closed multi-contour after we take into account
phase factors arising from non-trivial monodromies of the integrand.

To determine whether such a $\cC $ exists, it is convenient
to introduce new coordinates $(\ze_1,\ldots,\ze_n)$ on $M$ defined by
 $\ze_1=z_1$ and
$\ze_i=z_i/z_1$ for $i=2,\ldots,n$. This gives a decomposition of
$M$ into the product $\CC^*\times M'$, where $M'=\{ (\ze_2,\ldots,\ze_n)\in
\CC^{n-1} | \ze_i\neq0\,,\ze_i\neq1\,,\ze_i\neq \ze_j (i\neq j)\}$. The
corresponding factorization in homology (the Eilenberg-Zeiler theorem)
gives  $H_n(M,\cS) = H_1(\CC^*,\cS')\otimes
H_{n-1}(M',\cS'')$ (compare [\TK]).  Here $\cS'$ and  $\cS''$ are the induced
local systems determined by the change of variables, $z_i\rightarrow\ze_i$,
in the integral.
Since $H_1(\CC^*,\cS')$ is non-trivial only if $\cS'$ is constant,
which just means that the integrand is
single valued in $\ze_1$,  we find that  all terms in \PBCb\ vanish unless
$\La$  and $\be$ satisfy the following condition
\eqn\PBCd{
- \alp^2 \left( (\La+\rh ,\be) - \half (\be,\be) \right) \in\ZZ \,.
}

To summarize, we have shown that \PBCd, with $\be$ being a sum of
positive roots, gives a necessary condition for the operator
$Q_{\La,\La'}(\cC)$ to be a nontrivial intertwiner between two Fock
spaces.  In fact, this condition also appears to be sufficient. The
latter requires  a more detailed analysis of the integrals in \PBCb,
or, equivalently, of the homology classes $H_{n-1}(M',\cS'')$. For
$\widehat {sl(2)}_k$ it follows directly from the results in [\FF,\TK],
while arguments along these lines are given for the general case in
[\FFrd].

For a given $\La$, all $\be\in\De_+$ satisfying \PBCd\ can be determined quite
easily, after we rewrite \PBCd\  more concisely in terms of affine
roots and weights.  Let $l_1,\ldots,l_n$ be  integers for which the
corresponding residue in \PBCb\ is non-zero. Since the $L_0$ eigenvalue of
$\widehat{s}_{i,l}$ equals $-l$, it is natural to associate to
$\widehat{s}_{i,l
   }$ an
affine root $\widehat{\al}_{i,l} = l\de+ \al_i$, where $\de$ is the
(imaginary) root dual to $-L_0$. Thus we define
\eqn\PBCe{\eqalign{
\wLa  = \La + k\La_0\,&, \qquad  \wrh  = \rh + \dCo\La_0\,, \cr
\widehat{\al}_{i_k,l_k}  =  l_k\de + \al_{i_k}\,&, \qquad
\wbe = {\sum_k} \widehat{\al}_{i_k,l_k} \,.\cr}
}
Clearly,  $\wbe$  (which  is an  extension of  $\be$ to the affine root lattice
$\widehat P$) lies on the (half-)lattice
\eqn\PBCg{
\wbe\in\ZZ_+\cdot\left( \wDe_+^{(+)}\cup (-\wDe_+^{(-)})\right)\,,
}
where
\eqn\PBCh{
\eqalign{
\wDe_+^{(+)} & = \{ \wal = n\de + \al\, |\, \al\in\De_+\,, n\geq0\}\,,\cr
\wDe_+^{(-)} & = \{ \wal = n\de - \al\, |\, \al\in\De_+\,, n>0 \}\,. \cr}
}
In this notation \PBCd\ simply becomes
\eqn\PBCf{
(\wLa + \wrh, \wbe ) - \half (\wbe,\wbe) = 0\,,
}
which we recognize as the Kac-Kazhdan equation [\KK].

Recall that the Kac-Kazhdan equation \PBCf\ is the starting point for
the analysis of the decomposition structure of Verma modules, \ie of
the singular vector structure and the embedding pattern for the
corresponding Verma modules.  There, however,
$\wbe\in\ZZ_+\cdot\wDe_+$.  In the present case of Fock space modules,
as a consequence of \PBCg , some of the arrows in the embedding diagram
will get inverted, but the main structure nevertheless
remains the same.

In particular, setting $\wLa'=\wLa-\wbe$, equation
\PBCf\ leads to
\eqn\PBCi{
|\wLa' + \wrh |^2 = |\wLa + \wrh |^2\,,
}
implying
\eqn\PBCj{
\wLa' = w*\wLa \equiv w(\wLa+\wrh) - \wrh\,,
}
for some element  $w$ of the affine Weyl group $\wW$ of $\whg$. Projecting
this equation onto the weight lattice of $\bf g$, we  identify the set of
weights $\La'$ that satisfy \PBCd\ with the orbit of
$\La$ under the shifted action of $\widehat W$. [Recall that
if we represent $w=t_\ga w_0$,
where $w_0$ is an element of the Weyl group $W$ of
$\bf g$, $t_\ga$ is the translation by a vector $\ga$ in the long root
lattice of $\bfg$ and $\La$ is a weight of level $k$, then
$w*\La = w_0(\La+\rh)-\rh + (k+\dCo)\ga$.]

In view of \PBCj\ we are now left to consider the problem of determining
the existence of an intertwiner of
the form \PBCb\ between two Fock spaces $\cF_{w*\La}$ and $\cF_{w'*\La}$,
where
$w,w'\in\widehat W$.  In analogy with the similar analysis
in the case of Verma modules [\KK],  one would like to characterize the
solution
by some special  property of the corresponding elements of $\widehat W$.
Indeed,   \PBCg\ motivates the following definition of a ``twisted length''
on $\wW$ (see \eg [\FFrc,\BMPa])
% CUT
\def\procPBCk{\eqn\PBCk{{
\tilde{l}(w)  = | \Ph_w^{(+)} | - | \Ph_w^{(-)}|\,,\qquad
\Ph_w^{(\pm)} = \wDe_+^{(\pm)} \cap w( \wDe_-) \,.}}}

\procPBCk
%\prepPBCk
% ENDCUT
Further, writing $w\rightarrow w'$ if there exists an $\al\in\wDe_+$ such
that $w=r_\al w'$ and $\tilde{l}(w) = \tilde{l}(w')+1$, we may define
a partial ordering (``twisted Bruhat ordering'') by
$w\preceq w'$ iff there exist $w_1,\ldots,w_k\in\wW$ such that
$w\rightarrow w_1\rightarrow \ldots \rightarrow w_k
\rightarrow w'$. We find that
for integrable weights $\La$ of level $k$,  $\La\in P_+^k$,
\eqn\PBCl{
\fancyhatg (\cF_{w*\La}, \cF_{w'*\La}) \neq0
\qquad {\rm iff}\qquad w\preceq w'\,.
}

Explicit construction of the intertwiners \PBCb\ requires a more
detailed understanding of the closed multi-contours $\cC$.  Recently,
insight into the structure of $H_n(M,\cS)$ has been gained by the
observation that the space of (suitable) relative cycles carries a
representation of the quantum group $\cU_q({\bf g})$ at $q=\exp(\pi i
\alp^2)$, which is, in fact, isomorphic to a Verma module. The absolute
cycles correspond to the singular vectors in this Verma module
[\SVa--\FW]. The ``dual'' of this statement, which we will briefly
discuss below, was proved in [\BMPb] (see also [\GS,\RRR]).

Instead of varying the contour $\cC$ one may take a fixed contour and
instead vary the analytic continuation of the integrand. For the
contour $\cC$ we take a set of nested contours $|z_1|>\ldots>|z_n|$
(for $z_i\neq1$) where $z_i$ is integrated counterclockwise from $1$ to
$1$ around the origin $z=0$.  The ambiguity in the phase of the
integrand is fixed by analytic continuation from the region on the
positive real half axis where $0<z_n<\ldots<z_1$.  With these
conventions the operators $\nest{s_{i_1}\ldots s_{i_n}}$ and
$\nest{s_{\pi(i_1)}\ldots s_{\pi(i_n)}}$ related by a permutation $\pi$
are not necessarily proportional. We will show however that within
$\nest{\ldots}$ the $s_i$'s satisfy the Serre relations of
$\cU_q(\bfnm)$, thus clarifying, in this ``dual'' context, that the space
of potential intertwiners is isomorphic to a quantum group Verma
module.

Observe that upon interchanging two screening operators $s_{i}$ and
$s_j$ one picks up a phase factor $q^{(\al_i,\al_j)}$. This motivates
the following definition of the adjoint operator
\eqn\PBCaa{
({\rm ad}\ s_i) s_j = s_is_j - q^{(\al_i,\al_j)} s_js_i\,.
}
More generally
\eqn\PBCab{
({\rm ad\ }s_i) x = s_i x - q^{(\be,\al_i)}xs_i\,,
}
if $\be= \sum \al_{i_j}$ is the weight of the string of screening
currents $x= s_{i_1}\ldots s_{i_n}$.
With these notations the Serre relations for $s_i$ can be written as
[\Ro]
\eqn\PBCac{
({\rm ad\ }s_i)^{1-a_{ij}} s_j = 0\,,
}
where $a_{ij} = {2(\al_i,\al_j)\over (\al_i,\al_i)}$ is the Cartan
matrix of $\bfg$.
Now imagine writing  the integral over the contour $\cC$ to a sum of
integrals where all the variables are located on the unit circle and
have a specified ordering. Denote by $\cI_{i_1\ldots i_n}$ the integral
with ordering $0<{\rm arg\ }z_1<\ldots<{\rm arg\ }z_n<2\pi$ [\BMPb].
We claim that for arbitrary $N\geq1$
\eqn\PBCad{
({\rm ad\ }s_i)^N s_j = {\cal B}_{N} \cI_{i\ldots ij}\,,
}
where
\eqn\PBCae{
{\cal B}_N =
\prod_{\ka=0}^{N-1} \left( {(1-q^{\ka b_{ii} +2b_{ij}})(1-q^{(\ka+1)
b_{ii}})
\over 1-q^{b_{ii}}}\right) \,,\qquad b_{ij}\equiv (\al_i,\al_j)\,.
}
Clearly this implies the Serre relations \PBCac.\foot{We would like to
emphasize that the proof holds for arbitrary generalized Cartan matrices.}

Equation \PBCae\ is proved by induction. One easily checks the
assertion for $N=1$.
The induction step goes as follows
\def\adsi{{\rm ad\ }s_i}
\eqn\PBCaf{\eqalign{
(\adsi)& \left( (\adsi)^{N-1} s_j\right)  =
s_i \left( (\adsi)^{N-1}s_j\right) - q^{(N-1)b_{ii}+b_{ij}}
\left( (\adsi)^{N-1}s_j\right) s_i\cr
& = \left( {1-q^{Nb_{ii}}
\over 1-q^{b_{ii}}} \right) {\cal B}_{N-1} \cI_{i\ldots
ij} + q^{(N-1)b_{ii}+b_{ij}} {\cal B}_{N-1} \cI_{i\ldots iji} \cr
&\ \ \ - q^{(N-1)b_{ii}+b_{ij}} \left(
{\cal B}_{N-1} \cI_{i\ldots iji} +
\left( {1-q^{Nb_{ii}}\over 1-q^{b_{ii}}} \right)
{\cal B}_{N-1} \cI_{i\ldots ij}\right) \cr
& = {\cal B}_N \cI_{i\ldots ij}\,. \cr}}

It is now quite straightforward to determine which of the operators
\PBCa\ actually commute with $\whg_k$. This requires evaluating  the
commutator
 \eqn\PBCp{
\big[ \,f_i(z)\,,\,\nest{P(s)} \,\big] \,,
}
where $P(s)$ is some polynomial in screening operators.
Using \PBBi\  we find that the commutator consists of boundary terms at
${\rm arg\ }z_i=0$ and ${\rm arg\ }z_i=2\pi$ only. By carefully keeping
track of the phase factors that one picks up by crossing the variables
it can be established that the commutator \PBCp\ vanishes precisely
when the polynomial $P(s)$ corresponds to a singular vector in the
aforementioned quantum group Verma module. For details we refer to
[\BMPb]. It  is also clear that, in the language of homology with local
coefficients, the  vanishing of \PBCp\  is equivalent to the
corresponding contour $\cC$ being closed.  We conclude that the set of
intertwiners can be determined by  analysing the structure of singular
vectors in the corresponding quantum group Verma module.

The final step in the construction of the resolution is
to combine the set of intertwiners and Fock spaces
into a complex. We refer to [\BMPb,\BMPa] for
details.  This leads to
the following complex (\ie $d^{(n)}d^{(n-1)}=0$)
\eqn\PBCq{\matrix{
\ldots  & \mape{d^{(-2)}}& \cF_\La^{(-1)}& \mape{d^{(-1)}}& \cF_{\La}^{(0)}&
\mape{d^{(0)}}& \cF^{(1)}_\La& \mape{d^{(1)}}& \ldots\cr}
}
where for the case of $\La$ integrable
\eqn\JMCr{
\cF_\La^{(n)} = \bigoplus_{ \{ w\in\wW | \tilde{l}(w) = n\} }\
\cF_{ w*\La }\,.
}
This complex provides a resolution of the
irreducible module $L_\La$, \ie
\eqn\PBCs{
H^{(n)}_d(\cF_\La) = \cases{ L_\La & for $n=0$\cr
0 & otherwise $\,.$\cr}
}
The cohomology of \PBCq\ was computed rigorously for $\widehat
{sl(2)}_k$  [\BF,\FFrd]. In general case it is believed that a proof may
be given by showing, similarly as  in the case of finite-dimensional
Lie algebras,  an isomorphism between \PBCq\ and the so-called weak
resolution  of $L_\La$ constructed in [\FFrc].

For the purpose of obtaining the completely degenerate \hwm s of \Walg s
using the quantum Drinfeld-Sokolov reduction (Section 4) it will turn
out that considering only integrable weights is not sufficient.
Rather, we have to consider fractional levels, say
\eqn\PBCm{
k+\dCo = {p\over p'}\,,\quad{\rm gcd}(p,p')=1\,,\quad {\rm gcd}(p',r^\vee)
=1\,,\quad p\geq\dCo\,,\quad p'\geq{\rm h}\,,
}
and weights of the form
\eqn\PBCn{
\La = \Lap - (k+\dCo)\Lam\,,
}
with $\Lap\in P_+^{p-\dCo}$ and $\Lam\in P_+^{\vee\,p'-{\rm h}}$.
Here,
$r^\vee$ is the ``dual tier number'' of $\whg$. We recall that in the
simply laced case ${\rm h}=\dCo\,,P_+=P_+^\vee$ and $r^\vee=1$.

Weights of the form \PBCn\ are a subset of the class of so-called
admissible weights [\KWa]. In fact, for reasons that will become
clear in Sections 4 and 5, these weights are such that $\La- (k+\dCo)
\rh^\vee$
is a so-called nondegenerate principal admissible weight
[\KWc].\foot{Moreover, up to  Weyl reflections $w\in W$, these
exhaust the class of nondegenerate principal admissible weights.
For simplicity we ignore these Weyl reflected weights. They lead to
isomorphic \Walg\ modules anyhow.}

It turns out that the entire discussion of intertwiners between
Fock spaces $\cF_{w*\La}$, with $\La$  integrable, can easily be extended
to this more general class of admissible weights. For $\La$ of the form
\PBCn , condition \PBCd\ simply becomes
\eqn\PBXa{
-{p'\over (p-h^\vee)+h^\vee}\big((\La^{(+)}+\rho,\be)-\half(\be,\be)\big)
+(\La^{(-)},\be)\in\ZZ\,.}
However, since  $p$ and $p'$ are relatively prime and $\La^{(-)}$ is integral,
this is equivalent to \PBCd\ with $\La\rightarrow \La^{(+)}$ and
$k\rightarrow p-h^\vee$.
Similarly, the vanishing of \PBCp\ translates into the requirement
that $P(s)$ corresponds to a singular vector in the quantum group Verma
module of highest weight $\Lap$.
Thus, once more restricting to the intertwiners of the
form \PBCb, we  obtain immediately that
\eqn\PBCo{
\fancyhatg(\cF_{w*\Lap - (k+\dCo)\Lam},
\cF_{w'*\Lap - (k+\dCo)\Lam}) \neq0\qquad {\rm iff}\qquad w\preceq w'\,.
}
Moreover, using these homomorphisms we can build a complex as in \PBCq ,
but with
\eqn\PBCr{
\cF_\La^{(n)} = \bigoplus_{ \{ w\in\wW | \tilde{l}(w) = n\} }\
\cF_{ w*\Lap - (k+\dCo)\Lam }\,.
}
Since $d^{(i)}$ are precisely the same as in the resolution for the
integrable weight $\La^{(+)}$, we have indeed constructed a complex.
In fact we expect that this complex provides a Fock space resolution of
the irreducible module $L_\La$.  For $\widehat{sl(2)}$ this has been
proved in [\BF]. However, contrary to the integrable case, the
corresponding weak resolution for  arbitrary admissible representations
has not yet been constructed, so it is not clear how a general proof
should proceed.

One piece of evidence supporting the validity of \PBCs\  is that by
computing the character of $L_\La$, using the Euler-Poincar\'e
principle under the assumption that \PBCs\ holds, one recovers the
character formulae of Kac and Wakimoto [\KWa,\KWb].

In the subsequent sections we therefore assume the validity of
\PBCs\ and examine its consequences.

\newsec{Quantum Drinfeld-Sokolov reduction}

Consider the following 1-dimensional representation of $\whn\equiv
\bfnp\otimes\CC((t))$
\eqn\PBDa{
\ch_{DS}(e^\al (z) ) = \cases{ 1 & if $(\rh^\vee,\al)=1$,
\ie if $\al$ is a simple root\cr
0 & otherwise$\,.$ \cr}}
The quantum Drinfeld-Sokolov reduction amounts to enforcing
the constraint $e^\al(z)$ $\sim\ch_{DS}(e^\al(z))$ through a BRST procedure
(see \eg [\BO,\FFrb]).
The BRST operator corresponding to this constraint is given by
$Q= Q_0 + Q_1$, where
\eqn\PBDb{Q_0=  \oint {dz\over 2\pi i}
\left( \sum_{\al\in\De_+} c^\al(z) e^\al(z)
- \half \sum_{\al\be\ga\in\De_+} f^{\al\be\ga} :b^\ga(z)c^\al(z)c^\be(z):
\right)\,,}
is the standard differential associated to $\whn$, and
\eqn\PBDc{Q_1 =   - \oint {dz\over 2\pi i}
\sum_{\al\in\De_+} \left(c^\al(z)\ch_{DS}(e^\al(z))\right)\,.}
They satisfy
\eqn\PBDca{Q^2=Q_0^2=Q_1^2=\{ Q_0,Q_1\} = 0\,.}

Let $C^*(\whg_k,\whn)$ denote the completion of $\cU(\whg_k)\otimes
C\!\ell$ by infinite series ($C\!\ell$ is the Clifford algebra of the
fermionic ghost system). This space is graded by ghost number $({\rm gh})$.
We put $({\rm gh})(c^\al)=1=-({\rm gh})(b^\al)$.
The differential $Q$ acts on $C^*(\whg_k,\whn)$ by (super) commutation
and its cohomology, which is again a (graded) Lie algebra,
will be denoted by $\Heca{}$.  Roughly speaking this is just the
operator cohomology of Q restricted to the space of operators built from
the modes of the current algebra.  The algebra $\cW_k[\bfg]\equiv
\Heca{(0)}$ is called the \Walg\ associated to $\bfg$.  Indeed,
an examination of the spectral sequence associated to the double complex
with differentials $(Q_0,Q_1)$ shows that for generic $k$
(\ie $k\not\in -\dCo+\QQ_+$) the algebra $\cW{}_k[\bfg]$ can be identified
with the commutant of a set of screening charges
[\FFrb,\Fr], thus making contact with the definition of a \Walg\
given by Fateev and Lukyanov [\FLa,\FL].\foot{This is true for simply
laced $\bfg$. For non-simply laced the algebra $\cW[B_n]$ of Fateev-Lukyanov,
for example,
rather corresponds to $\cW[B(0,n)]$ in the terminology of this paper
(see \eg [\Wat,\It]).}
It has been conjectured that $\Heca{(n)}\simeq0$ for $n\neq0$ [\FFrb,\Fr].

The Drinfeld-Sokolov reduction procedure clearly at the same time provides
the representation spaces for the symmetry algebra.  For, by taking
the cohomology of $Q$ on modules $M\otimes \cF^{bc}$,
where $M$ is any $\whg_k$-module from the category $\cO$ [\DGK,\RW],
we obtain a functor
sending $\whg_k$-modules $M\in\cO$ to modules $\Hecm{}{M}$ of the
``Hecke algebra'' $\Heca{}$. In particular, $\Hecm{(n)}{M}$ becomes
a $\Heca{(0)}$-module for each $n\in\ZZ$.

The algebra $\cW_k[\bfg]$ will, at least, contain the Virasoro algebra.
An explicit representative is given by
\eqn\PBDd{
T(z) = T^{Sug}(z) + \rh^\vee\cdot\p h(z) + T^{gh}(z)\,,
}
where $T^{Sug}$ is the conventional Sugawara operator, and
\eqn\PBDe{
T^{gh}(z) = \sum_{\al\in\De_+} \left( ((\rh^\vee,\al)-1)
:b^\al\p c^\al:(z)
+ (\rh^\vee,\al) :\p b^\al c^\al:(z) \right)\,.
}
Of course, the ``improvement terms'' in \PBDd\ just amount to requiring
that the BRST current $j(z)= j_0(z)+j_1(z)$ becomes a primary field of
conformal dimension one  so that $Q$ and $T(z)$ commute.  The central
charge of this Virasoro algebra is
\eqn\PBDf{\eqalign{
c & = {k\  {\rm dim\ } {\bf g} \over k+\dCo } - 12k |\rh^\vee|^2
 - 2\sum \left( 6(\rh^\vee,\al)^2 - 6(\rh^\vee,\al) +1 \right) \cr
& = \ell - 12|\alp\rh + \al_-\rh^\vee|^2\,, \cr}}
where we have introduced $\al_+\al_-=-1$,
and ${\alp}^{-2} = k+\dCo$, as before.
The conformal dimension of a $\whg_k$-module $M$ with highest weight
$\La$ under \PBDd\ becomes
\eqn\PBDg{
h_\La = \half\alp^2(\La,\La+2\rh) - (\La,\rh^\vee) = \half
(\La\al_+,\La\al_+ + 2(\alp\rh+\al_-\rh^\vee))\,.
}
For simply laced Lie algebras $\bfg$,
levels $k$ as in \PBCm\ and weights $\La$ as in \PBCn, the
equations \PBDf\ and \PBDg\ can be recognized as the central
charges and conformal dimensions of the completely degenerate \hwm s
of the \Walg\ minimal models [\FL,\BBSS] (remember $\rh=\rh^\vee$, and
identify $\al_0=\alp+\al_-$).\foot{In the case of $\cW$-algebras
it is conventional to parametrize \hwm s through $\La\al_+=
\Lap\al_+ + \Lam\al_-$.}
This, ultimately, is the explanation of our choice \PBCn.
Note that \PBDg\ can be written as $\half\alp^2(\La',\La'+2\rh)$ where
$\La' = \La - (k+\dCo)\rh^\vee$. Thus \PBDg\ can be interpreted as the
(untwisted) conformal dimension of the nondegerate principal
admissible weight $\La'$.

Now suppose we take $M=L_\La$. From the above it is reasonable
to expect that, for weights
$\La$ as in \PBCn, we have $\Hecm{(n)}{L_\La}\simeq0$ for $n\neq0$
while $\Hecm{(0)}{L_\La}$ $\simeq L_\La^\cW$, where $L_\La^\cW$ is an
irreducible \hwm\ of the algebra $\cW_k[\bfg]$.\foot{For weights $\La$
not of the form \PBCn\ this is not necessarily true. One can show, for
instance, that for dominant integer weights $\Hecm{(n)}{L_\La}\simeq0$
for all $n\in\ZZ$ [\FFrd,\FFrb]. Also, the fact
that $\Hecm{(0)}{L_\La}$ is an irreducible \Walg\ \hwm\ is not
a priori obvious but nevertheless appears to be true for weights of the
form \PBCn.}
We will provide some evidence in support of this conjecture in Section 5.

Assuming then that $\Hecm{(n)}{L_\La}\simeq L_\La^\cW \de^{n,0}$, there
exists a natural candidate for a resolution of $L_\La^\cW$, obtained
simply by applying the functor $\Hecm{}{\,\cdot\,}$ to a resolution
of the $\whg_k$-module
$L_\La$. So suppose, following Section 3, that we are given a resolution of
the $\whg_k$-module $L_\La$ where the differential $d$ is constructed out
of the screening operators $s_i(z)$ and the terms are free field Fock spaces
$\cF^{(n)}_\La \equiv \cF^{(n)\ph}_\La\otimes \cF^{\be\ga}$. Then,
using the fact that $[\,Q,d\,]=0$, as well as the result
(see \eg [\BO--\FFrb])
\eqn\PBDh{
\Hecm{(n)}{ \cF^\ph_\La\otimes\cF^{\be\ga}}
\simeq \de^{n,0}\,\cF^\ph_\La \,,
}
we have
\eqn\PBDha{\eqalign{
L_\La^\cW & \simeq \Hecm{}{L_\La} \simeq \Hecm{}{H_d(\cF^{(*)\ph}_\La\otimes
\cF^{\be\ga})}\cr &  \simeq H_d( \Hecm{}{\cF^{(*)\ph}_\La\otimes
\cF^{\be\ga}}) \simeq H_d(\cF^{(*)\ph}_\La)\,.\cr}
}
That is, we have constructed a resolution of the \Walg\ module
$L_\La^\cW$ in terms of free field Fock spaces $\cF^{(n)\ph}_\La$.

In fact, one can show that in $Q$-cohomology [\BO,\Fi]
\eqn\PBDi{
s_i(z) \ \sim\ \ :e^{-i\al_+\al_i\cdot\ph(z)}:\ \  \equiv \tilde{s}_i(z)\,,
}
as well as
\eqn\PBDia{
T(z) \ \sim\  -\half :\p\ph(z)\cdot\p\ph(z): -
(\alp\rh + \al_-\rh^\vee)\cdot i \p^2\ph(z) \,.
}
Equation \PBDi\ implies that the differential $d$ of the resolution
\PBDha\ can be replaced by the differential $\tilde{d}$ obtained from
$d$ by replacing the screening operators $s_i(z)$ by their
$Q$-cohomologous counterparts $\tilde{s}_i(z)$ in terms of scalar
fields $\ph(z)$ only. Furthermore, since the \Walg\ is exactly the commutant
of the screening charges $\oint \tilde{s}_i(z)$ as remarked before,
this confirms that the \Walg\ commutes with $\tilde{d}$ and thus acts on
all the terms in the resolution. For $\whg = \widehat{sl(2)}$ the
differential $\tilde{d}$ can be recognized as the differential
of Felder's complex that provides a resolution of the Virasoro
degenerate \hwm s [\Fe]. This observation constitutes a proof
of \PBDha\ for $\widehat{sl(2)}$, as has been discussed in [\BO].

Equation \PBDia\ gives an explicit expression for the Virasoro part
of this commutant and confirms equations \PBDf\ and \PBDg.
For $\bfg = A_n$ one can find a generating function for the other
generators $W_k(z)$ of the \Walg\ by means of the so-called quantum
Miura transformation [\FLa,\FL]
\eqn\PBDj{
\sum_{k=0}^{n+1} W_k(z) (\al_0\p)^{n+1-k} =\,
:\! (\al_0\p - \ep_1\cdot i\p\ph(z))(\al_0\p - \ep_2\cdot i\p\ph(z))\cdots
(\al_0\p - \ep_{n+1}\cdot i\p\ph(z))\!:
}
where $\{ \ep_i\,,i=,\ldots,n+1\}$
is the set of weights of the vector representation of $A_n$, normalized
according to $\ep_i\cdot\ep_j = \de_{ij} - {1\over n+1}$, and such that
the simple roots of $A_n$ are given by $\al_i = \ep_i - \ep_{i+1}$.

\newsec{Some character formulae}

In this section we apply the Euler-Poincar\'e principle to the resolution
of Section~4, to derive a character formula for the (completely
degenerate) irreducible \hwm s $L^\cW_\La$. We compare the result to the
formulae obtained from the coset approach to \Walg s [\BBSS].
For notational simplicity we restrict ourselves to simply laced Lie
algebras $\bfg$.

So, we take $\La= \Lap - (k+\dCo) \Lam$ as in \PBCn. Recall
\eqn\PBEa{
{c\over 24}  = {1\over 24} (\ell - 12\al_0^2|\rh|^2) ={\ell\over24} -
{(p-p')^2\over 2pp'} |\rh|^2\,,
}
\eqn\PBEaa{
h_{w*\Lap - (k+\dCo)\Lam} - {c\over24}
=  -{\ell\over24} + {1\over2pp'} |p' w(\Lap+\rh) - p(\Lam+\rh)|^2\,.
}
Under the assumption $\Hecm{}{L_\La}\simeq L^\cW_\La$ the character
${\rm ch\,}_{L_\La^\cW}$ of
$L_\La^\cW$ can be calculated using \PBEaa\  and the
Euler-Poincar\'e principle (as usual $q=\exp(2\pi i\ta)$)
\foot{For this character to be nonvanishing
it is essential that the admissible weight $\La'\equiv \La - (k+\dCo)\rh$
be nondegenerate, \ie that there do not exist roots $\al\in\De_+$ such
that $(\La',\al)\in\ZZ_+$.}
\eqn\PBEb{
\eqalign{
{\rm ch\,}_{L_\La^\cW}(\ta) & = \Tr _{L_\La^\cW}\,q^{L_0 - c/24} =
\sum_{n\in\ZZ} (-1)^n\Tr_{H^{(n)}_Q(L_\La)}\,q^{L_0-c/24} \cr
& = \sum_{n\in\ZZ} (-1)^n \Tr_{H_d^{(n)}(\cF^\ph_\La)}\,q^{L_0-c/24} =
\sum_{n\in\ZZ} (-1)^n\Tr_{\cF^{(n)\ph}_\La}\, q^{L_0-c/24}  \cr
& =  {1\over\et(\ta)^\ell }
\sum_{w\in\widehat{W}} \ep(w) q^{ \bigoneoverp| p' w(\Lap+\rh) -
p(\Lam+\rh)|^2} \,.\cr}
}

We now briefly compare the above to the results that follow from the
coset approach to \Walg s. We recall that for $\La\in P^{1}_+$ we have
a $\Wg$ algebra realized on $L_\La\otimes M$ for any $\whg$-module $M$ from
the category $\cO$ (see [\BBSS] for $\cW[A_2]$). The generators of $\Wg$
can be expressed in terms of the generators of $\whg_1\oplus\whg_k$.
This $\Wg$-algebra
commutes with the diagonal action of $\whg_{k+1}$.

In particular, taking $M=L_{\La'}$ with principal admissible weight
$\wLa'=\wLa^{(+)} - (u-1)(k+\dCo)\La_0\,, \wLa^{(+)}\in P_+^{p-\dCo}$,
where we have parametrized $k+\dCo = p/(p'-p) = p/u$ leads to the following
branching function for the occurrence  of $L_{\La''}\,, \wLa'' =
\wLa^{(-)} - (u-1)(k+\dCo+1)\La_0\,, \wLa^{(-)}\in P_+^{p'-\dCo}$ in
the decomposition of $L_{\La}\otimes L_{\La'}$ under the diagonal
$\whg_{k+1}$ action [\KWa,\KWc]
\eqn\PBEd{
b_{\La\otimes\La'}^{\La''} = {1\over\et(\ta)^\ell} \sum_{w\in\wW}
\ep(w) q^{ \bigoneoverp |p'w(\Lap +\rh) - p(\Lam+\rh)|^2 }\,.
}
This formula is in agreement with \PBEb. For the unitary case
$p'=p+1$ formula \PBEd\ was derived in [\BMPc] from a free field
resolution for coset models.

The agreement of \PBEb\ and \PBEd\ can be considered not only as an
indication of the correctness of the assumption \PBDha , but at the same
time as an indication that the \Walg s derived from the coset $\whg_1\oplus
\whg_{k-\dCo} / \whg_{k-\dCo+1}$ and the quantum
Drinfeld-Sokolov reduction of $\whg_k$ are,
in fact, isomorphic.

\acknow

\footatend\immediate\closeout\rfile\writestoppt
\baselineskip=14pt{\bigskip\noindent {\bf  7. References}}%
\bigskip{\frenchspacing%
\parindent=20pt\escapechar=` \input refs.tmp\vfill\eject}\nonfrenchspacing

\vfill\eject\end